# Rapid laser-induced photochemical conversion of sol-gel precursors to $In_2O_3$ layers and their application in thin-film transistors

Spilios Dellis[1], Ivan Isakov[2], Nikolaos Kalfagiannis[1], Kornelius Tetzner[2], Thomas D. Anthopoulos[2], Demosthenes C. Koutsogeorgis*[1]

[1] Department of Physics, School of Science and Technology, Nottingham Trent University, Nottingham, NG11 8NS, UK.
[2] Department of Physics and Centre for Plastic Electronics, Imperial College London, London, SW7 2AZ, UK.

**Abstract:** We report the development of indium oxide ($In_2O_3$) transistors via a single step laser-induced photochemical conversion process of a sol-gel metal oxide precursor. Through careful optimization of the laser annealing conditions we demonstrated successful conversion of the precursor to $In_2O_3$ and its subsequent implementation in n-channel transistors with electron mobility up to 13 $cm^2$/Vs. Importantly, the process does not require thermal annealing making it compatible with temperature sensitive materials such as plastic. On the other hand, the spatial conversion/densification of the sol-gel layer eliminates additional process steps associated with semiconductor patterning and hence significantly reduces fabrication complexity and cost. Our work demonstrates unambiguously that laser-induced photochemical conversion of sol-gel metal oxide precursors can be rapid and compatible with large-area electronics manufacturing.

**Keywords:** $In_2O_3$, Laser Annealing, Transparent Thin Film Transistor, Sol-Gel, Metal Oxide Semiconductors

Thin-film transistors (TFTs) based on transparent metal oxide semiconductors hold great promise for a variety of emerging applications in large-area/volume electronics, including, but not limited to, flexible displays,[1] radio frequency identification (RFID) tags,[2] and transparent electronic devices.[3] To this end, recent years have witnessed the development of a wide range of high-mobility metal oxide semiconductors and devices that can be manufactured over large areas employing simple fabrication methods.[4] Among the various deposition techniques demonstrated, solution processing offers a scalable and cost effective route for high throughput and large-area deposition of various oxide materials including ZnO, $In_2O_3$, $In_4ZnO$, and InGaZnO (IGZO).[5,6]

Despite the huge promise, however, accurate control over the morphology and the chemical composition of solution-grown metal-oxides still remains challenging, leading to significant device-to-device performance variations. In addition, deposition of semiconducting metal oxides by "sol-gel" has so far been limited to high processing temperatures, typically in excess of >350 °C, rendering the technology incompatible with inexpensive, temperature-sensitive substrates such as plastic, the material of choice for large-scale roll-to-roll (R2R) processes.[7–9] This post-deposition heat treatment is a major obstacle that hinders the integration of heat-sensitive flexible polymeric substrates. Therefore, it is essential to develop alternative methods of processing that can meet the demands relevant to this area. Thermal annealing at reduced processing temperature, ideally less than 150 °C, has been examined aiming at oxide TFTs fabrication on flexible polymeric substrates.[10] However, the processing times that are typically required are very long (up to 4 h) and the achieved TFTs performance and stability are largely inadequate for real electronic applications.[11]

Optical annealing can be viewed as a powerful tool towards low-temperature fabrication schemes, however, recent reports on optical sintering suffer from lengthy exposure that renders the process unsuitable for high throughput R2R manufacturing.[9] Therefore, alternative methods that can deliver rapid and scalable materials processing are urgently required.[12] To this end, Laser Annealing (LA) offers fast processing along with rapid, precise and selective energy delivery in area and depth via critical laser energy absorption.[13,14] For instance, it has been shown that the use of ultrashort ultraviolet (UV) laser pulses can result in a significant enhancement of the electronic properties of the semiconductor, which in many cases cannot be achieved with conventional annealing.[15]

Despite the tremendous potential, however, to date only few studies have reported the use of LA for the fabrication/post-processing of metal oxide thin films. Imai et al.[16] showed that a high number of low fluence laser pulses (up to 1800 pulses of 20 $mJ/cm^2$) produced from an ArF excimer laser (193 nm) were able to remove organic compounds and induce crystallization in numerous initially amorphous metal oxide films. Low-fluence/multiple-pulse LA with XeCl excimer laser (308 nm) was suggested by Yang et al. in the fabrication of IGZO TFTs resulting in a field-effect mobility of 7.65 $cm^2$/Vs.[8] A different approach was recommended by Tsay et al. where the researchers employed high fluence and a low number of pulses (15) from a KrF excimer laser (248 nm) for improving the physical properties of IGZO thin film.[17]

$In_2O_3$ is a wide band-gap semiconducting material with great perspectives in the transparent electronics technology as it exhibits both high optical transparency and superior electronic properties with electron mobility of up to 220 $cm^2$/Vs reported for single crystals.[18–21] For this reason, a number of studies on the fabrication of $In_2O_3$ TFTs using high-temperature thermal annealing or a combination of thermal and optical annealing have been reported in recent years.[9,12,22,23] Here, we describe the fabrication of fully functional and high performance n-channel $In_2O_3$ TFTs, via LA in ambient atmosphere. During the laser-induced photo-conversion process, the sample temperature never exceeds the 150 °C, making the proposed approach compatible with plastic substrates. Despite the low thermal budget, the resulting TFTs exhibit excellent operating characteristics with electron mobility values of up to 13 $cm^2/Vs$.

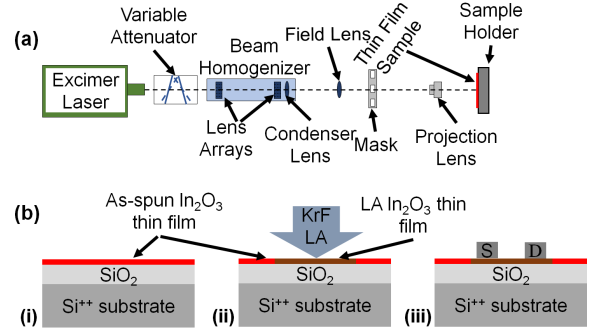

**Figure 1.** (a) The LA experimental setup used in the present work. (b) Schematic representation of the fabrication process of BG-TC $In_2O_3$ TFTs on Si/$SiO_2$ substrates.

Doped silicon (Si-n$^{++}$) wafers with a thermally grown 400 nm-thick layer of $SiO_2$ acting as gate and gate dielectric respectively, were used for the fabrication of bottom-gate, top-contact transistors (BG-TC). Prior to the deposition of the $In_2O_3$ layer, the substrates were cleaned by ultrasonication in deionized water, acetone, and isopropanol, with each step lasting 10 min. Substrates were subsequently exposed to atmospheric pressure UV ozone treatment for 10 min at room temperature. The metal oxide precursor solution consisted of indium nitrate and 2-methoxyethanol with a concentration of 20 mg/ml and was spun-coated directly on the Si$^{++}$/$SiO_2$ substrates for 30 s at 4000 rpm (Figure 1b (i)). As-spun precursor layers were thermally stabilized by drying them on a hot-plate at 150 °C for 15 min in order to remove excessive solvent residues.[24] LA (Figure 1b (ii)) was carried out with a KrF Excimer laser (LAMBDA PHYSIK LPX 305i), which is capable of delivering 25 ns pulses of unpolarized light at 248 nm with energy up to 1200 mJ per pulse. The beam delivery system (Figure 1a) comprises a variable attenuator, a beam homogenizer and a mask projection system. An XYZ translational stage was used in order to manipulate the sample during the laser processing. The laser spot delivered onto the samples was set to be a 2.5 × 2.5 $mm^2$ spot, with fluence (energy density) uniformity better than 2% throughout the spot. An extensive experimental schedule was followed, varying the number of pulses (1 to 10 at 1 Hz repetition rate) and the fluence (100–450 mJ/$cm^2$ with a step of 50 mJ/$cm^2$). After the LA step, TFT fabrication was completed with the thermal evaporation of the aluminum source and drain electrodes in high vacuum (<$10^{-5}$ mbar) using shadow masks (Figure 1b (iii)). The width and length of the TFT channels were 11 mm and 100 µm, respectively. TFT characterization was carried out in a nitrogen atmosphere (samples maintained at room temperature) using an Agilent B2902A parameter analyzer. The field-effect electron mobility was extracted from the linear operating regime using the gradual channel approximation according to:

$$\mu = \frac{L}{W C_i V_D} \frac{\partial I_D}{\partial V_G} \quad (1)$$

where, $W$ is the channel width, $L$ the channel length, $I_D$ the channel current in linear operating regime, $V_G$ the gate voltage, $V_D$ the drain-source voltage and $C_i$ the geometric capacitance of the gate dielectric. The as-spun $In_2O_3$ (after stabilization at 150 °C for 15 min) exhibits an insulating character that remained unaffected even after applying up to 10 pulses of fluence lower than 300 mJ/$cm^2$. Figure 2 displays the transfer characteristics of LA $In_2O_3$ TFTs prepared with 10 pulses at 300 mJ/$cm^2$. The device exhibits electron transport (n-channel) behavior with threshold voltage ($V_{th}$) of approximately –60 V and onset voltage ($V_{ON}$) below -100 V (minimum range measured). For the devices fabricated with fluence up to 400 mJ/$cm^2$, $V_{th}$ is in the range of –25 to –120 V and remains relatively constant with higher number of pulses (Figure 3a). Above 450 mJ/$cm^2$, however, $V_{th}$ progressively shifts to more negative values, turning the film severely conductive at higher number of pulses. Figure 3b shows the field-effect mobility measured in the linear regime for all LA $In_2O_3$ TFTs. All the devices exhibit high mobility with typical values in the range 9-13 $cm^2$/Vs. These results highlight the tremendous potential of LA as an alternative processing route for the development of metal oxide TFTs.

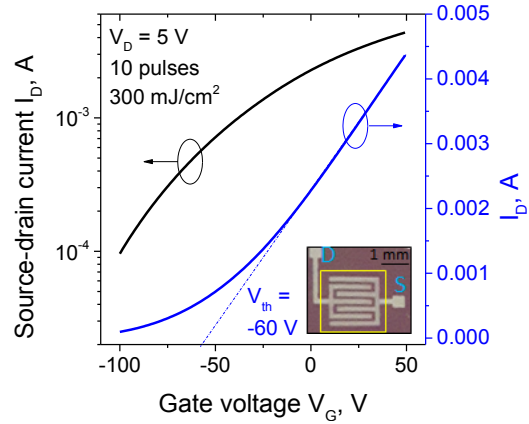

**Figure 2.** Transfer characteristics of the LA $In_2O_3$ TFTs (10 pulses at 300 mJ/cm$^2$); inset: optical micrograph of a complete device showing drain and source contacts and outline of the area subjected to LA.

A potential drawback of LA of $In_2O_3$ is the large negative $V_{th}$ observed in all TFTs, as it can lead to high power consumption. To address this an additional process step was implemented in order to shift $V_{th}$ towards $V_G = 0$ V. The step comprises of a mild thermal annealing of the TFTs following complete fabrication (to include source and drain electrodes), during which the samples were heated to 100 °C for 60 min. Post-annealed devices showed $V_{th}$ much closer to zero, especially for devices below 400 mJ/cm$^2$, albeit accompanied with a slight reduction in electron mobility, as shown in Figure 3a and 3b respectively. The transfer and output characteristics of the LA $In_2O_3$ TFT (10 pulse at 300 mJ/cm$^2$, the same device shown in Figure 2) with mild post-fabrication thermal annealing are presented in Figure 4. As can be seen, the $V_{th}$ and $V_{ON}$ shift to +5 V and –60 V, respectively. In this case the $V_G$ range was sufficient for studying the TFT performance in both on and off-state and allowed the $I_{ON}/I_{OFF}$ to be calculated as higher than 10$^6$. This further thermal treatment step is not expected to affect the $I_{ON}/I_{OFF}$ but only to shift the $V_{th}$ to more positive values. The µ is found to marginally decrease, probably due to the device not reaching its maximum mobility at the gate-voltages used. These trends are representative for all LA $In_2O_3$ TFTs with average $I_{ON}/I_{OFF}$ ratio of >10$^5$, $V_{th}$ close to zero, and µ in the range of 6-10 cm$^2$/Vs. Output characteristics shown in Figure 4b indicate Ohmic-like injecting behavior, negligible gate leakage in the low voltage regime and saturation regime at high voltages.

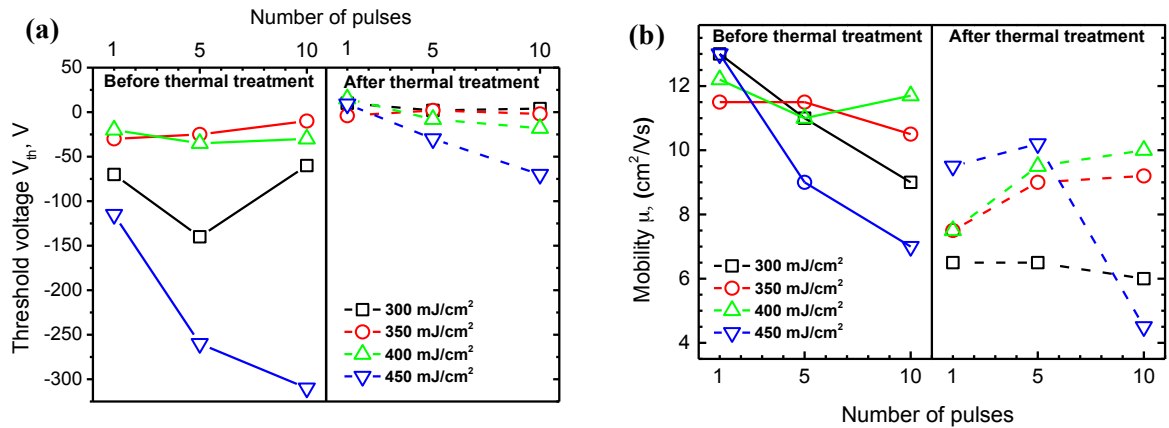

**Figure 3.** (a) Threshold voltage and (b) mobility of the LA $In_2O_3$ TFTs before (solid lines) and after (dashed lines) the post fabrication mild-thermal treatment at 100 °C for different LA processing. The lines in this figure serve only as a guide to the eyes.

$In_2O_3$ TFT devices were also prepared by thermal annealing at 250 °C for 60 min, as a reference for comparison (Figure 4a, dashed lines). Although the $V_{ON}$ of the devices and their sub-threshold slope were superior to the LA ones, the µ values were significantly lower and in the order of 0.4 cm$^2$/Vs. This relatively low µ value as compared to previously published data,[24] could be the result of the prolonged exposure of $In_2O_3$ to non-controlled environment; all samples were exposed to ambient air for 2 weeks during LA of the $In_2O_3$ layers, S-D deposition and electrical characterization.[9] Therefore, an overall device process optimization is expected to lead to significantly improved TFT performance. Despite the nonidealities, the average μ obtained for LA $In_2O_3$ TFTs is very high if one considers the low thermal budget and very high speed of the process.

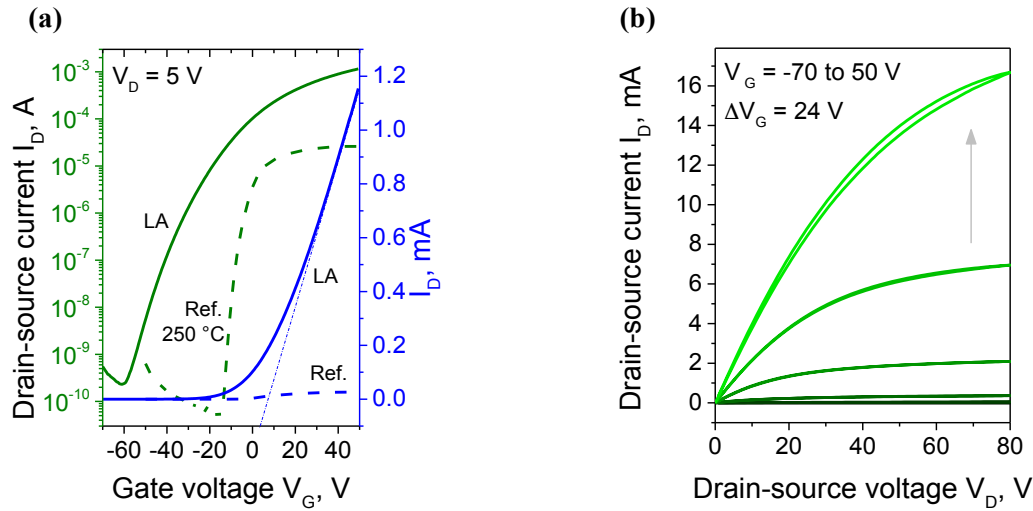

**Figure 4**. (a) Transfer characteristics of the LA $In_2O_3$ TFT (10 pulses at 300 mJ/cm$^2$) after the mild-thermal treatment. For comparison the transfer characteristics of the reference device are also shown (dashed lines); (b) output curves of the same device.

Another advantage of the LA process is the removal of the necessity of any $In_2O_3$ patterning step. The latter is a common practice for all metal oxide TFTs fabricated by conventional non-additive deposition techniques, including vacuum and solution-phase processes, in order to electrically isolate the individual device.[22] Unlike traditional methods, LA can be performed on-demand on selected areas of the substrate. Upon LA, only the irradiated area of the precursor film is converted from a soluble insulating state to an insoluble metal oxide semiconductor, leaving the rest of the sol-gel film unaffected. As a consequence, the resulting TFTs exhibit very low gate leakage current that is typically around $10^{-10}$ A (our measurement limit). Such low leakage current is critical for the successful implementation of any TFT technology in commercial electronic products. Furthermore, the unconverted regions of the sol-gel layer can be easily removed by immersing the entire substrate to a suitable solvent bath. Although not relevant to this work, such simplified direct patterning process could well be exploited in the future for the manufacturing of complementary circuitry involving patterning of p- and n-channel TFTs in close proximity to each other on the same substrate, as a substitution for an elaborate and hence costly conventional patterning process.

In summary, we demonstrated the application of laser annealing for the rapid chemical conversion of a metal oxide sol-gel precursor to a high electron mobility semiconductor and its subsequent implementation in n-channel $In_2O_3$ TFTs. Key characteristics of the LA process include: (i) solution deposition of the precursor sol-gel material, (ii) high speed chemical conversion of the precursor to the semiconducting state via controlled LA, and (iii) low thermal budget that does not exceed 150 °C during any stage of the fabrication process. Additionally, the localized and selective character of the LA process allows for the spatial conversion of the precursor layer to semiconductor, leaving the unexposed sol-gel layer in its insulating state. This leads to TFTs with extremely low gate leakage current, an attribute that can only be otherwise achieved through the introduction of additional, and hence costly, semiconductor patterning steps. Finally, the proposed LA process does not rely on controlled atmosphere, hence further simplifying its application and potential scale up.

**Acknowledgements:**

This work was conducted under the Pathfinder project LAFLEXEL, that was funded by the Centre of Innovative Manufacturing for Large Area Electronics (CIMLAE) and hence in extension by the EPSRC. K.T. and T.D.A. also acknowledge financial support from the People Programme (Marie Curie Actions) of the European Union's Framework Programme Horizon2020: "Flexible Complementary Hybrid Integrated Circuits" (FlexCHIC), grant agreement no. 658563.

[1] J.K. Jeong, Semicond. Sci. Technol. **26**, 34008 (2011).
[2] A. Nathan, A. Ahnood, M.T. Cole, S. Lee, Y. Suzuki, P. Hiralal, F. Bonaccorso, T. Hasan, L. Garcia-Gancedo, A. Dyadyusha, S. Haque, P. Andrew, S. Hofmann, J. Moultrie, D. Chu, A.J. Flewitt, A.C. Ferrari, M.J. Kelly, J. Robertson, G.A.J. Amaratunga, and W.I. Milne, Proc. IEEE **100**, 1486 (2012).
[3] L. Petti, N. Munzenrieder, C. Vogt, H. Faber, L. Buthe, G. Cantarella, F. Bottacchi, T.D. Anthopoulos, and G. Troster, Appl. Phys. Rev. **21303**, 1 (2016).

[4] Y. Sun and J.A. Rogers, Adv. Mater. **19**, 1897 (2007).
[5] P.F. Moonen, I. Yakimets, and J. Huskens, Adv. Mater. **24**, 5526 (2012).
[6] S.R. Thomas, P. Pattanasattayavong, and T.D. Anthopoulos, Chem. Soc. Rev. **42**, 6910 (2013).
[7] K.K. Banger, R.L. Peterson, K. Mori, Y. Yamashita, T. Leedham, and H. Sirringhaus, Chem. Mater. **26**, 1195 (2014).
[8] Y.-H. Yang, S.S. Yang, and K.-S. Chou, IEEE Electron Device Lett. **31**, 969 (2010).
[9] Y.-H. Kim, J.-S. Heo, T.-H. Kim, S.K. Park, M.-H. Yoon, J. Kim, M.S. Oh, G.-R. Yi, Y.-Y. Noh, and S.K. Park, Nature **489**, 128 (2012).
[10] S. Yang, J.Y. Bak, S. Yoon, M.K. Ryu, H. Oh, C. Hwang, G.H. Kim, S.K. Park, and J. Jang, IEEE Electron Device Lett. **32**, 1692 (2011).
[11] G.J. Lee, J. Kim, J.-H. Kim, S.M. Jeong, J.E. Jang, and J. Jeong, Semicond. Sci. Technol. **29**, 35003 (2014).
[12] J. Leppäniemi, K. Ojanperä, T. Kololuoma, O.-H. Huttunen, J. Dahl, M. Tuominen, P. Laukkanen, H. Majumdar, and A. Alastalo, Appl. Phys. Lett. **105**, 113514 (2014).
[13] N. Kalfagiannis, A. Siozios, D. V. Bellas, D. Toliopoulos, L. Bowen, N. Pliatsikas, W.M. Cranton, C. Kosmidis, D.C. Koutsogeorgis, E. Lidorikis, and P. Patsalas, Nanoscale **8**, 8236 (2016).
[14] C. Tsakonas, W. Cranton, F. Li, K. Abusabee, A. Flewitt, D. Koutsogeorgis, and R. Ranson, J. Phys. D. Appl. Phys. **46**, 95305 (2013).
[15] M. Nakata, K. Takechi, T. Eguchi, E. Tokumitsu, H. Yamaguchi, and S. Kaneko, Jpn. J. Appl. Phys. **48**, 81608 (2009).
[16] H. Imai, A. Tominaga, H. Hirashima, M. Toki, and N. Asakuma, J. Appl. Phys. **85**, 203 (1999).
[17] C.Y. Tsay and T.T. Huang, Mater. Chem. Phys. **140**, 365 (2013).
[18] Z. Galazka, R. Uecker, and R. Fornari, J. Cryst. Growth **388**, 61 (2014).
[19] O. Bierwagen and J.S. Speck, Appl. Phys. Lett. **97**, 97 (2010).
[20] R.L. Weiher, J. Appl. Phys. **33**, 2834 (1962).
[21] K. Nomura, H. Ohta, A. Takagi, T. Kamiya, M. Hirano, and H. Hosono, Nature **432**, 488 (2004).
[22] H. Faber, Y.-H. Lin, S.R. Thomas, K. Zhao, N. Pliatsikas, M.A. McLachlan, A. Amassian, P.A. Patsalas, and T.D. Anthopoulos, ACS Appl. Mater. Interfaces **7**, 782 (2015).
[23] L. Petti, H. Faber, N. Münzenrieder, G. Cantarella, P.A. Patsalas, G. Tröster, and T.D. Anthopoulos, Appl. Phys. Lett. **106**, 92105 (2015).
[24] Y.-H. Lin, H. Faber, J.G. Labram, E. Stratakis, L. Sygellou, E. Kymakis, N. a. Hastas, R. Li, K. Zhao, A. Amassian, N.D. Treat, M. McLachlan, and T.D. Anthopoulos, Adv. Sci. **2**, 1500058 (2015).